\documentclass[conference]{IEEEtran}
\IEEEoverridecommandlockouts

\usepackage{comment}
\usepackage{xspace}
\usepackage{url}
\usepackage{cite}
\usepackage{graphicx}
\usepackage{listings}
\usepackage{xcolor}
\usepackage[inline]{enumitem}
\usepackage{flushend}

\newcommand{\ompi}{\textsc{Open~MPI}\xspace}
\newcommand{\ulfm}{\textsc{ULFM}\xspace}

\newcommand{\mpicode}[1]{\texttt{\detokenize{#1}}\xspace}

\lstdefinelanguage[MPI]{C}[]{C}{keywordsprefix={MPI_}}

\begin{document}

\title{Implicit Actions and Non-blocking Failure Recovery with MPI}

\author{
\IEEEauthorblockN{Aurelien Bouteiller}
\IEEEauthorblockA{Innovative Computing Laboratory\\The University of Tennessee\\
Knoxville, Tennessee, USA\\
Orcid: 0000-0001-5108-509X\\Email: bouteill@icl.utk.edu}
\and
\IEEEauthorblockN{George Bosilca}
\IEEEauthorblockA{Innovative Computing Laboratory\\The University of Tennessee\\
Knoxville, Tennessee, USA\\
Orcid: 0000-0003-2411-8495\\Email: bosilca@icl.utk.edu}
}

\maketitle

\begin{abstract}
Scientific applications have long embraced the MPI as the environment of choice to execute on large distributed systems.
The User-Level Failure Mitigation (ULFM) specification extends the MPI standard to address resilience and enable MPI applications to restore their communication capability after a failure.
This works builds upon the wide body of experience gained in the field to
eliminate a gap between current practice and the ideal, more asynchronous, recovery model in which the fault tolerance activities of multiple components can be carried out simultaneously and overlap.
This work proposes to:
(1) provide the required consistency in fault reporting to applications (i.e., enable an application to assess the success of a computational phase without incurring an unacceptable performance hit);
(2) bring forward the building blocks that permit the effective scoping of fault recovery in an application, so that independent components in an application can recover without interfering with each other, and separate groups of processes in the application can recover independently or in unison; and
(3) overlap recovery activities necessary to restore the consistency of the system (e.g., eviction of faulty processes from the communication group) with application recovery activities (e.g., dataset restoration from checkpoints).
\end{abstract}


\section{Introduction}\label{sec:intro}

Numerical simulations, data mining, and---most recently---machine learning have taken on critical roles in all fields of scientific exploration.
High-performance computing (HPC) systems are specifically designed to provide the cutting-edge computing capabilities required to obtain timely answers to the crucial scientific challenges of our time, and---as such---are routinely breaking new barriers in harnessing a massive number of computing resources to deliver staggering levels of performance.
Applications have widely embraced the Message Passing Interface (MPI)~\cite{mpi3} to cope with the challenges posed by the deployment of scientific computation on distributed systems.
That MPI is a standardized and stable specification, and that all HPC vendors provide an implementation tailored to deliver performance on their systems, ensures that users are confident that their application will easily port and achieve consistently high performance on current and future hardware.

Unfortunately, MPI lacks a comprehensive management system for failures, and resilience (or lack thereof) could still pose a
significant challenge for next-generation HPC systems~\cite{2012-12,2013-5}.
Since the Petascale era, reliability is already measured in days (e.g., production runs on Oak Ridge National Laboratory's (ORNL's) Titan system experience close to three failures per day~\cite{Gamell:2014}).
The mean time between failures (MTBF) is dependent on the number of components in the system, and, despite expected major advances in the reliability of individual hardware components, and the prevalence of ``fat'' GPU nodes in currently deployed exascale systems (putting us in the optimistic part of pre-exascale predictions~\cite{IESP}), the MTBF for post-exascale systems could still decline significantly, or at the minimum, act as an external constraint that restricts machine design and component choices (and prices) to those that do can sustain an acceptable MTBF.

The HPC community has already started reacting to this threat and is actively investigating resilient applications and mitigation techniques that permit the efficient execution of scientific computing on machines that may experience multiple fault events per allocation.
To accompany this community effort, we have designed the User-Level Failure Mitigation (ULFM) extension to the MPI specification~\cite{DBLP:journals/ijhpca/BlandBHBD13} and have provided a reference implementation~\cite{Bland:2012tp}. 
The ULFM specification provides the basic infrastructure to
1) detect and report failures of MPI processes;
2) interrupt the flow of the application;
and 3) repair the state of MPI communication objects to restore the full communication capability (e.g., the ability to carry out collective communication).
This early effort has clearly responded to a need in the community (see Section~\ref{sec:related} and~\cite{LOSADA2020467}), with multiple international research teams sustaining a two-pronged exploratory effort to investigate algorithmic fault tolerance techniques in fields as varied as physics, chemistry, and engineering on one hand, and a flurry of fault tolerance frameworks designed to ease and streamline the expression of resilience techniques on the other.

This paper explores how to again broaden the realm of possible methodologies for resilience techniques by exploring asynchronous and scoped techniques for error reporting and consistency of state recovery in the message passing layer.
The overarching goal is to enable the concurrent recovery of multiple components of the software stack.
Two major statements motivate this effort.
First, the cost of recovering the message passing's full communication capability can be significant, and---in some instances---in the same order of magnitude as the cost of recovering the application state.
Given the appropriate level of asynchrony, these complementary activities could overlap, thus significantly driving down the overall cost of recovery.
Second, applications are typically composed of multiple software components in charge of different software functional modules.
These modules can recover part of the application state independently, in a completely scoped insulation from the rest of the application, but may also require knowledge about a broader scope in order to trigger the recovery path at the appropriate moment and in the appropriate order to enable all modules to recover successfully.
To expound further, this work explores how to:
\begin{enumerate*}[label=(\arabic*\upshape)]%
\item provide the required consistency in fault reporting to applications, that is, enable an application to assess the potential success of a computational phase without incurring an unacceptable performance hit;
\item bring forward the building blocks that permit the effective scoping of fault recovery in an application so that independent components in an application can recover without interfering with each other, and separate groups of processes in the application can recover independently or in unison, as needed;
\item overlap recovery activities that are necessary to restore the consistency of the system (e.g., eviction of faulty processes from the communication group) with application recovery activities (e.g., dataset restoration from checkpoints); and
\item deliver these novel capabilities in a mature software environment (Open MPI) to ensure an easy path forward for application communities
interested in fault tolerance research.
\end{enumerate*}

The rest of this paper is organized as follows: in Section~\ref{sec:bg} presents the background for MPI failure recovery and general \ulfm concepts;
Section~\ref{sec:prop:scope} presents a technique to control the scope, that is, the set of processes and MPI operations, that get notified when a failure occurs;
Section~\ref{sec:prop:uniform} discusses a new way of reporting errors in collective operations that simplifies a coordinated response to the failure;
Section~\ref{sec:prop:ishrink} presents a non-blocking MPI state rebuilding routing that enables more asynchrony and ease ordering constraints during recovery;
Section~\ref{sec:perf} presents the experimental results evaluating these concepts;
Section~\ref{sec:related} presents related works for MPI recovery systems, as well as discuss the applicability of the proposed ideas to a large set of fault tolerant applications and frameworks that currently use \ulfm;
and finally Section~\ref{sec:conclusion} concludes.

\section{Background}\label{sec:bg}

\begin{figure}
	\includegraphics[width=\linewidth]{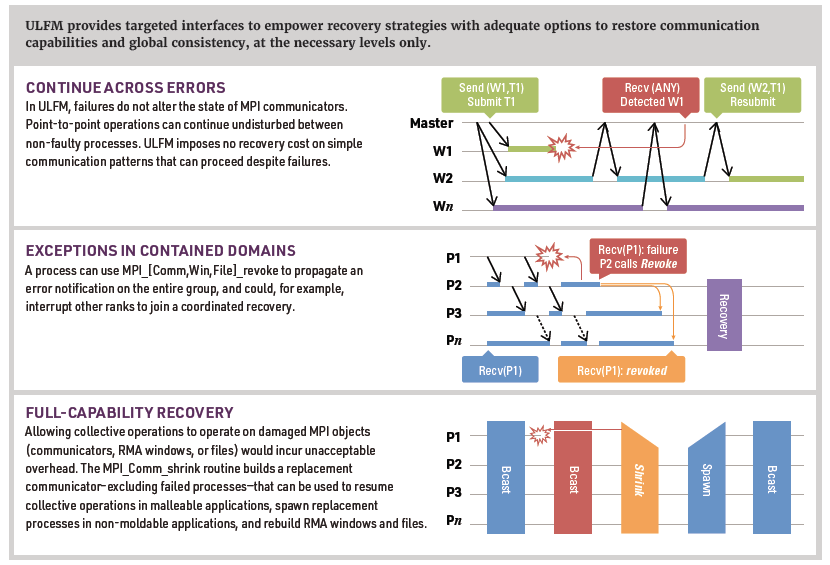}
	\caption{Major use-cases for the ULFM MPI fault tolerance extension.}
	\label{fig:ulfm1}
\end{figure}

\begin{figure*}
	\includegraphics[width=\linewidth]{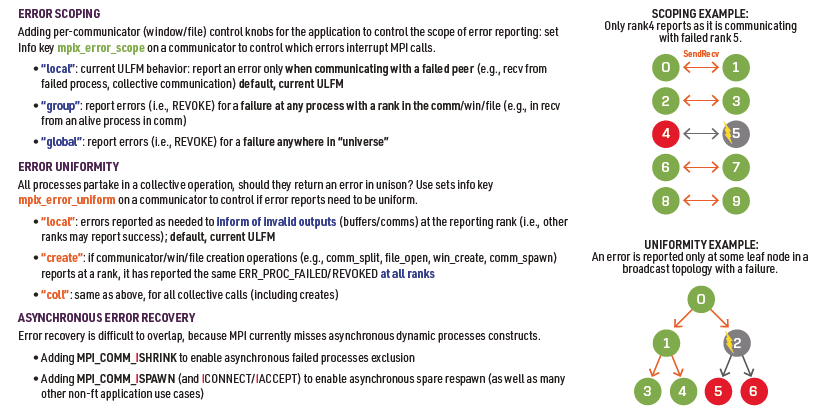}
	\caption{New uniformity and scope modes in relation to existing \ulfm definitions.}
	\label{fig:error_scoping1}
\end{figure*}

Research in fault management techniques has a long and exceptionally rich history~\cite{cappello2009fault}.
In some instances, there is a possibility that a hardware failure could result in drastic, life or death consequences.
In mission-critical applications, when safety trumps performance, solutions such as duplication, triplication, were a long time favorite, as they permit detecting and mitigating faults at a high but predictable production and/or operation cost.
In contrast, HPC considerations often call for a less drastic skew in the tradeoff between safety and performance.
Although scientific computation requires a high level of confidence in the accuracy of the result produced, hardware failures (i.e., when a failure results in some processes of the application terminating unexpectedly) often manifest by preventing the completion of the application rather than corrupting the end results.
Even for soft errors (i.e., when a failure results in a corruption of the computation accuracy or correctness), the problem is often reframed as an efficiency tradeoff, with the additional constraint that detecting the error may require periodic and extensive data invariant verification~\cite{FASI2016522,ftcg} or an increase in per-operation cost to provide online error detection~\cite{Li:2013:RAF:2503210.2503226}.
But once the failure has been asserted, recovery techniques are often very similar to those deployed in crash-failure scenarios, with restarting from a known good state (checkpoint) often being an adequate strategy~\cite{bosilca2015composing,7284417}.
Thus, what users are generally seeking in HPC is the sweet spot where the cost of fault mitigation is lower than the potential downtime, delay, and duplicate computation cost incurred by the crash of the application~\cite{DBLP:journals/concurrency/BosilcaBBCDGHRVZ14}.
Historically, global checkpointing~\cite{2007-1,BHKLC06,cl85,SSBLDHR03} has been the mitigation technique of choice.
However, performance models predict that we be may on the verge of an era where---despite its relative simplicity---more advanced techniques (checkpoint based or otherwise) will yield better efficiency~\cite{fthpl2011,Bouteiller:2015:AFT:2737841.2686892,DBLP:journals/concurrency/BosilcaBBCDGHRVZ14}.
This fact has steered a significant effort from the HPC community in exploring tailored techniques for tolerating failures in applications on one hand, and in exploring abstractions, runtime systems, and programming models that can effectively support the execution of fault tolerant applications on the other hand.

The \ulfm specification~\cite{DBLP:journals/ijhpca/BlandBHBD13} has recently taken a central role in this community (see Section~\ref{sec:related}).
It permits the continued execution of MPI programs after a failure by providing three critical capabilities:
(1) the detection of failures and reporting of errors;
(2) the ability to reliably disseminate a signal to interrupt the communication flow of the application;
and (3) a consensus operation that permits stabilizing the state of the application and resuming normal operation after MPI internal structures and capabilities have been restored.
In more details, following the diagrams in Figure~\ref{fig:ulfm1}, we can follow \ulfm operating workflow by inspecting use-cases.
Some applications can run-through faults, simply continuing to execute a point-to-point communication pattern that avoids communicating with known failed processes; typical for the class are manager-worker frameworks that simply resubmit the work delegated to a dead worker to another compute resource.
For such applications, it is important that the effect of failure is the most localized, and that operation between non-failed processes continues normally.
Some applications are malleable, that is, they require the full capabilities of MPI after recovery (e.g., collective communication) but can redistribute the work to execute on a smaller set of processes; such applications can use the \mpicode{MPIX_COMM_SHRINK} operation to recreate working communicators that exclude dead processes.
Finally, some applications have an inflexible mapping that requires an exact number of processes; the ULFM specification can be used in combination with MPI-2 dynamic operations, such as \mpicode{MPI_COMM_SPAWN} to recreate an isomorphic communicator.
Applications that are not run-through will often require a collective recovery procedure where all processes join to repair the MPI state, and the application dataset.
New and original algorithms were developed to ensure the scalability of the resilience features in~\ulfm.
In~\cite{DBLP:conf/pvm/BouteillerBD15}, we proposed a scalable and reliable broadcast algorithm to efficiently support the \emph{revoke} operation.
In \cite{DBLP:conf/sc/HeraultBBGTPD15}, we proposed an original consensus algorithm tailored for HPC systems.
In~\cite{DBLP:conf/sc/BosilcaBGHRSD16}, we designed a scalable failure detector, and in~\cite{DBLP:conf/pvm/HoriYHBBI15}, we studied replacement processes.
With these developments, interest swelled among MPI implementors~\cite{DBLP:conf/ccgrid/BlandLSB15} and scientific communities as they experimented with these capabilities to design fault tolerant frameworks and programming language extensions that support failure recovery, as well as algorithmic resilience methods over a range of applications.
Feedback from early adopters has identified potential areas of improvement that we have set to explore in this paper.

\section{A Novel Approach to MPI Recovery}
\label{sec:prop}
In the following subsections, we will describe the concepts we introduce to improve the reactivity, expressivity, and potential for overlap of recovery activities in MPI programs.

\subsection{Error Scoping}
\label{sec:prop:scope}

Failure detection and subsequent error reporting is the starting point
of the entire recovery procedure. In \ulfm, the scope of error reporting is
limited to processes that perform a direct communication operation with a failed
process (Figure \ref{fig:error_scoping1}). The rationale is twofold. First, some application use cases
 operate under the assumption that as long as a
 can complete successfully (i.e., the group of processes participating
 in the operation does not contain a failed process) it must proceed
 without interruption. This approach is particularly tailored for job-stealing
 types of applications, where the failure of some workers is irrelevant
 to the operation of the remaining workers as well as to controller processes
 that do not manage these workers. Second, general failure
detectors~\cite{Chandra:1996p5068,1630918,vanRenesse98,SWIM} can generate a significant
number of background control messages, either when they require the
establishment of pervasive heartbeat monitoring, or when employing
gossip techniques. To preserve performance, the original
\ulfm specification was designed to enable an implementation where
only in-band error detection is provided (i.e., errors collected
directly from the transport layer of MPI when the connection to an
endpoint is broken at the network driver level).

However, prior results hint that: (1) the cost of out-of-band failure
detection can be minimal and scalable when implemented correctly,
as demonstrated by the performance of the failure-detection algorithm
we designed for \ulfm~\cite{DBLP:journals/ijhpca/BosilcaBGHRSD18}, and (2) when
considering complex applications comprised on multiple software modules,
certain modules may require more thorough information about errors that
happen outside their direct neighbors in the communication pattern of that
module; for example, it may be required for this component to be informed
of any failure happening at any process on which this module spans, regardless
of whether or not the module is communicating with this process at the moment.

The first aspect is to define the interface with which the MPI user can
control the scoping behavior of the implementation to meet the desired
criterion. Again, we added a control for the error-scoping behavior
on a \emph{per-communicator} basis, with the addition of MPI info keys that
permit setting the specific desired behavior. That way application
modules that require a different error scoping behavior
can operate independently in different modes by posting the communication on
different communicators. The modes we envision will permit: (1) reporting only
errors when the posted communication operation cannot be completed, that is
one of the peers participating in the communication has failed, and the
resultant non-compliant behavior of the operation needs to be reported (this
represents the current \ulfm mode); (2) reporting an error when any process
in the communicator is detected as failed, even if the failed process is not
participating in reporting the MPI communication; and (3) reporting an error when
any process failure in the MPI universe is detected.
Given that the group
and global reporting mode require a more stringent failure detector that
can obtain out-of-band failure information, we identify the need to
thoroughly investigate and compare the cost of these operating modes with
the in-band only (i.e., from the network driver) detection cost.

One motivation for the group and global scope of error
reporting is to ease the writing of fault tolerant applications.
In many applications with a collective recovery pattern, \ulfm users
have to deploy a two-stage recovery process that first explicitly
propagates the occurrence of a fault event (using the \mpicode{MPIX_COMM_REVOKE}
operation) and only then proceeds to the collective recovery pattern.
This explicit management is very flexible, but we expect the scoped
recovery to significantly improve on the complexity of the error management
code by streamlining multiple error paths (either an error is directly reported, or
it has been triggered from an explicit propagation) into unified
management.

\begin{figure}
    \centering
	\includegraphics[width=.7\linewidth]{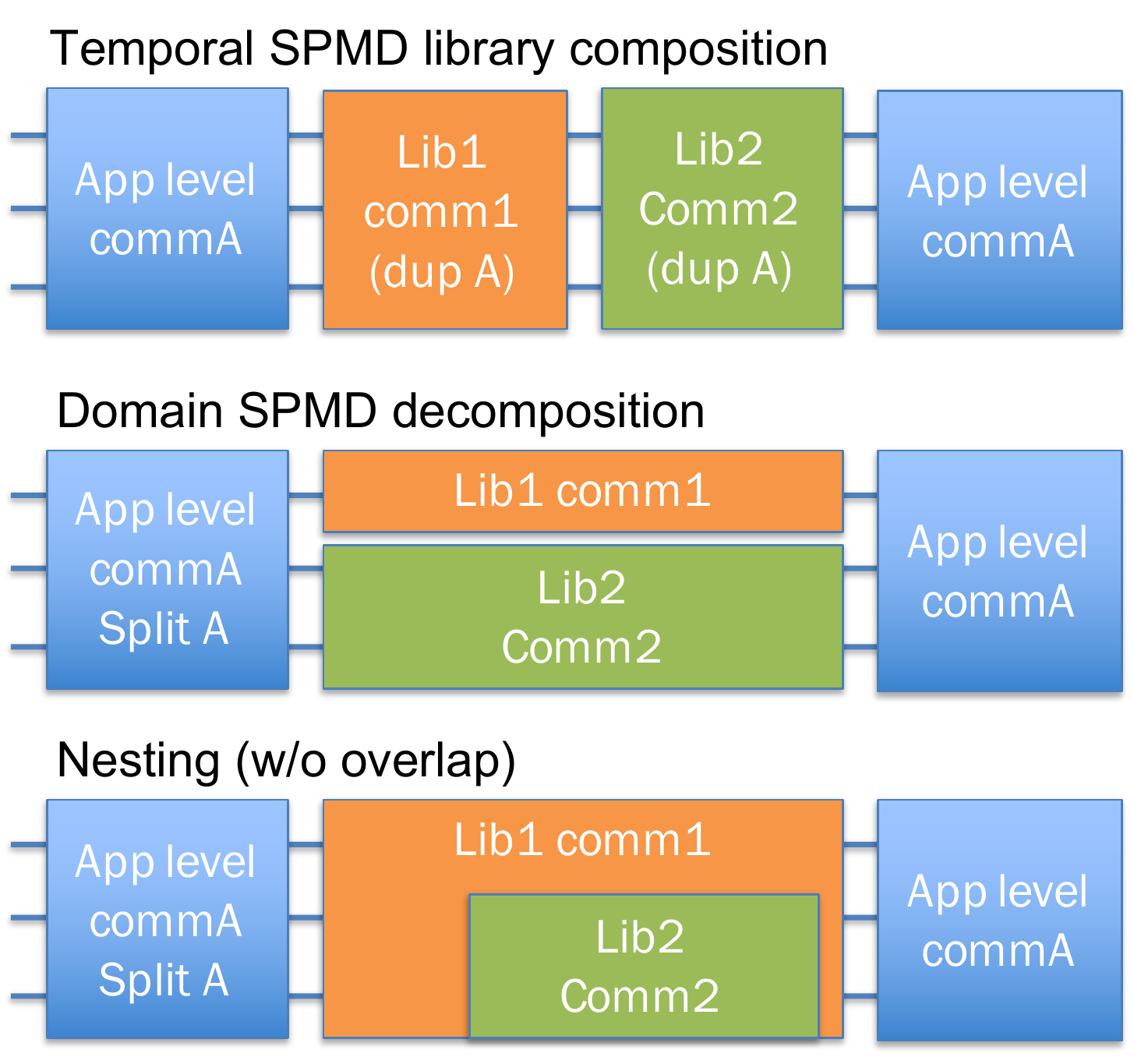}
	\caption{Applications with multiple libraries may have processes using communicator handles that are not known by one of the libraries, calling for a method to observe faults at processes that are not part of the local communication pattern.}
	\label{fig:libs}
\end{figure}

Another critical motivation for the global scoping mode is enabling the
cooperation of independent application modules or libraries that operate in the
application. Although modules generally operate independently and
require their communication to be segregated (i.e., employ separate
communicators), many recovery techniques are global in nature, and a fault
must trigger a recovery action in all modules of the application---including in processes that
may not be currently involved in communications  for the module.
Consider for example the case of a rank-domain decomposition (Fig.~\ref{fig:libs}) where the application subdivides some ranks to work with library 1, while the rest of the ranks work with library 2.
If the application requires a coordinated recovery action, both library modules must be interrupted; however, the library 1 does not have access to the handles for the communicators of library 2, and cannot directly revoke them.
A solution is for both libraries to observe the global scope, which will interrupt implicitly both libraries when a fault occurs in any of the domains.
Listing~\ref{code:lib1} demonstrates how to setup the library communicator in such a mode so that the library can detect faults at processes that are not currently engaged with the library-specific communicator (i.e., they may communicate in a separate communicator when the fault manifests).

\begin{lstlisting}[language={[MPI]C}, caption={Example of a library observing faults at processes that are not currently communicating on the library-specific communicator.}, label=code:lib1]
int lib1_init(MPI_Comm world) {
  /* ... */
  MPI_Comm_rank(world, &rank);
  MPI_Comm_size(world, &size);
  /* create a subcommunicator with only 1/4
   * the processes */
  int color = (rank < size/4)? 1: MPI_UNDEFINED;
  MPI_Comm_split(world, color, rank, &subcomm);
  if(MPI_UNDEFINED != color) {
    MPI_Info_create(&info);
    MPI_Info_set(info, "mpix_error_range", "universe");
    MPI_Comm_set_info(subcomm, info);
    /* From now on, a failure at a process that is **not**
     * calling lib1 will still cause communications on
     * subcomm to report errors */
    /* ... */
\end{lstlisting}

\subsection{Error Uniformity}
\label{sec:prop:uniform}

Collective communications in MPI are a powerful tool for expressing patterns
where a group of processes participate in the same communication
operation. By abstracting the intent of the communication pattern, the
operation can then be optimized by the MPI implementation using communication
topologies that limit the number of active connections between peers and
improve the operation's latency, and
pipelining techniques that improve the achievable bandwidth.
Collective operations also help
structure the code of the application: operations like the
\mpicode{MPI_BARRIER} are specifically indented for separating
algorithmic epochs.
A simplistic assumption postulates that such structuring features
would translate intact into a regime where failures may disrupt the
execution of the application. To understand why this usually
entails an unacceptable overhead, consider the case of
a typical implementation of the \mpicode{MPI_BCAST} operation over
a tree topology. In the normal, fault-free operation, as processes
relay the message from the root toward the leaf processes, they can
complete the broadcast operation successfully as soon as they have
forwarded the contribution. Thus, if a descendant process were to fail before
forwarding the broadcast message, its parents would not return an error, yet any
process in the subtree rooted
at the failed process would have to do so, since the message never reached them.
Note that this problem is not specific to one-to-all patterns, but can
happen symmetrically in all-to-one (and all-to-all) patterns.
Thus, in order to return an error at all ranks \emph{uniformly}, the
processes would have to \emph{agree} on the outcome of the call, which is an
extra fault tolerant synchronization step that is not required by the
semantic of the underlying communication. Thus, to protect
fault-free performance, in \ulfm, the collective communication errors are
defined as non-uniform (i.e., processes may return a different error
status for the same operation), and the \mpicode{MPI_COMM_AGREE} operation
provides an explicit means for users to synchronize in a fault tolerant fashion
when the need arises.

We identify the potential for simultaneously improving the
performance and simplifying the expression of common usage patterns.
First, some MPI collective operations often require a strong validation when
deployed in practice. For example, operations that create new communicators
often need to be validated for global success, or global failure, immediately.
Similarly, in many iterative algorithms, the reduction step that computes
the termination criterion is an excellent verification and restart point
for the application, leading to many users' employing an \emph{allreduce}
operation immediately followed by an \emph{agreement}. Thus, in this work, we
propose to provide the user with a level of control on the uniformity of
error reporting. We propose setting a \emph{per-communicator} property (practically,
by setting specific MPI Info keys on the communicator) that
lets the user decide whether collective operations on the communicator
should operate at maximum speed (non-uniform) or with implicit safety
(uniform error reporting at all ranks). This control enables a
distinction between pure communication operations (which are often critical for
performance) and setup/management operations (like operations that create
new communicators, or operations that change the logical epoch in the algorithm).

Our implementation of the concept adds an
agreement to all collective operations (an operation that can be performed
in approximately 2$\times$ the latency of a small message, non fault--tolerant
\emph{allreduce}~\cite{DBLP:conf/sc/HeraultBBGTPD15}).


\subsection{Asynchronous Recovery}
\label{sec:prop:ishrink}

Many applications need to restore the full communication capability of
MPI before they can resume their computational activity after a failure.
The \mpicode{MPI_COMM_SHRINK} operation provides the operational
construct that permits recreating a fully functional communication
context in which not only select point-to-point communication but
also collective communication can be carried. The core of the operation
relies on agreeing on the set of failed processes that need to be excluded
from the input communicator and then producing a resultant communicator with
a well-specified membership of processes. The current definition of
the shrink operation is blocking and synchronous, which limits the
opportunity for overlapping its cost.

We introduce a new non-blocking variant of the shrink
operation, \mpicode{MPIX_COMM_ISHRINK(comm, newcomm, req)}. To support
this operation, we designed a non-blocking variant of the
consensus-like elimination of failed processes as well
as the selection of the internal context identifier (a unique number
that is used to match MPI messages to the correct communicator on which
the operation was posted).

The expected advantages are multiple. Some applications may have to recover
multiple communicators after a failure. A common approach is to shrink the largest
communicator (e.g., \mpicode{MPI_COMM_WORLD}), and then recreate the damaged communicators with non-fault
 tolerant constructs (e.g., \mpicode{MPI_COMM_DUP}, \mpicode{MPI_COMM_SPLIT}, etc.), and then validate the whole
 set of new communicators with \mpicode{MPI_COMM_AGREE} on the large communicator.
 This is an effective approach when the derived communicators have the same,
 or similar number of processes in their respective groups. This can however be
 sub-optimal in applications that create small neighbor communicators with orders of
 magnitude fewer processes in their respective groups than in the overarching communicator.
 In such a situation, directly shrinking multiple time smaller communicators can be advantageous.
 With the availability of the non-blocking shrink, these multiple shrink operation have an
 opportunity to overlap one another.

The non-blocking shrink also gives an opportunity for overlap between the
cost of rebuilding communicators with the cost of restoring the application
dataset (e.g., reloading checkpoints, computing checksum on data, etc.).



\begin{figure}
    \includegraphics[width=\linewidth]{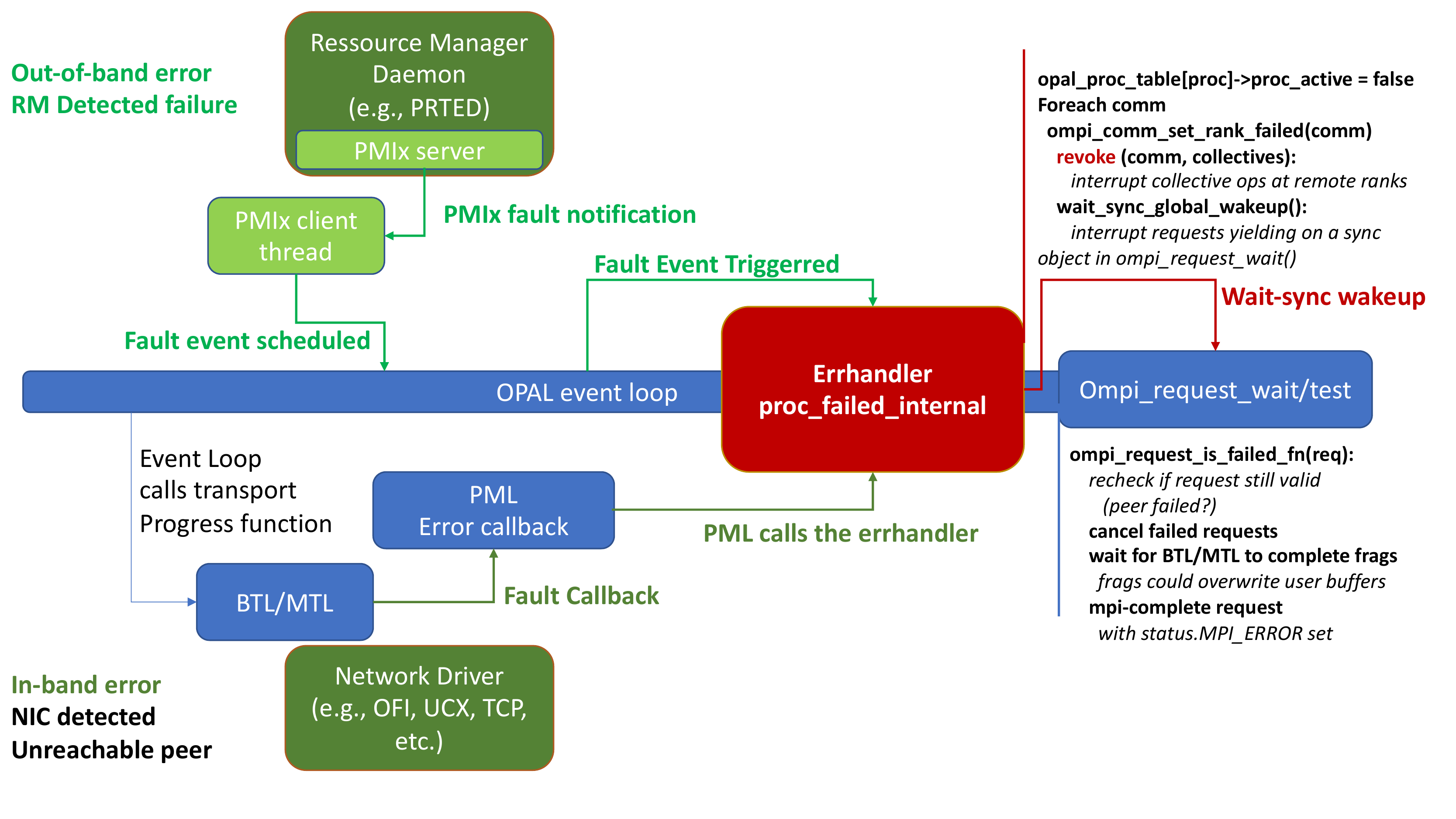}
    \caption{Implementation of Out-of-band error reporting in Open MPI}
    \label{fig:oob-faults}
\end{figure}

\begin{figure*}
    \includegraphics[width=\linewidth]{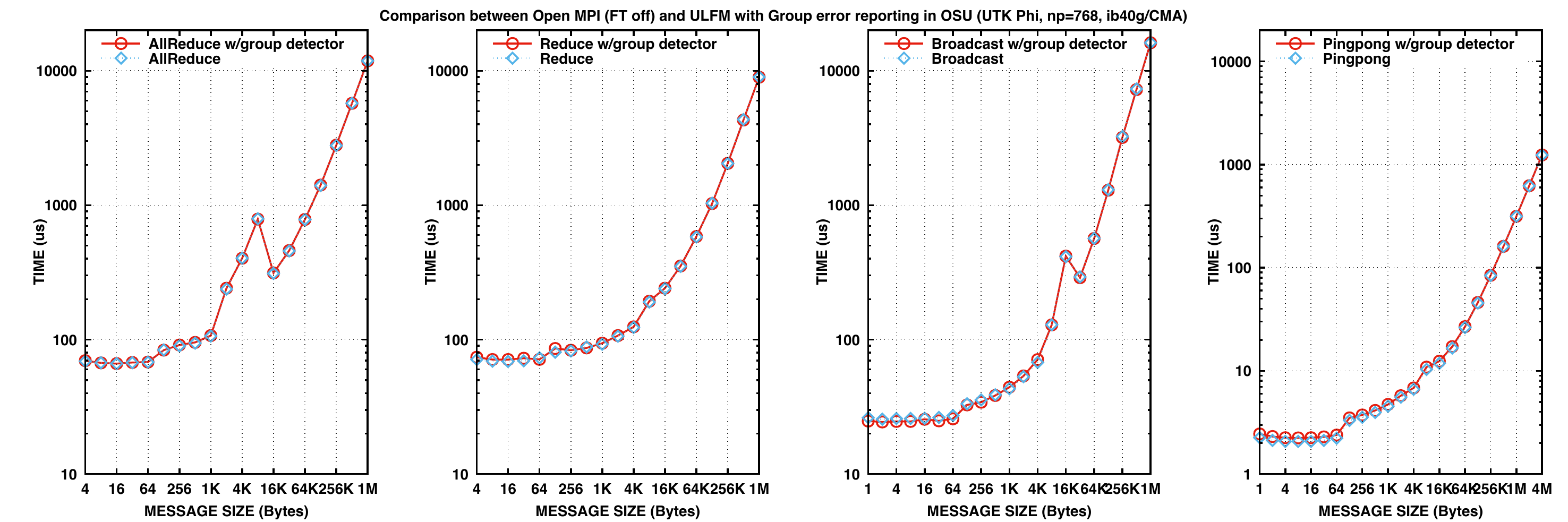}
    \caption{Performance comparison between in-band only (Local scope) and out-of-band (Group scope) failure detection (768 MPI ranks)}
    \label{fig:osuE}
\end{figure*}

\subsection{Implementation}
\label{prop:implementation}

These three ideas are implemented in the \ulfm reference implementation, which is currently integrated into the main \ompi development branch and slated for distribution with \ompi version 5.

A major new requirement with the introduction of the error reporting modes is that the implementation may not only use in-band error reporting.
In-band error reporting (bottom pathway in Fig.~\ref{fig:oob-faults} is when the network driver itself reports fault events, that get propagated through the software stack (usually through function return codes, and then in the internal status of the MPI request), until they bubble to the locus of the main MPI procedure call that completes the associated MPI operation.
Because the group and global error reporting modes have to produce errors when peers for which there are no active communication with fail, a purely in-band error reporting strategy is insufficient, and the implementation must contain an out-of-band mechanism to observe, and disseminate fault knowledge so that appropriate MPI errors can be produced.
In our \ompi implementation, the out-of-band error detection capability is delegated to the PRTE runtime (PMIx pathway in Fig.~\ref{fig:oob-faults}.) Faults are then bounced from the PRTE runtime event loop to the MPI event loop, and then produced to the user from MPI procedures on communicators that have the appropriate error reporting mode set.
The impact of running an out-of-band failure detector is one of the discussion we focus on in the performance evaluation section.

\begin{figure*}
    \includegraphics[width=\linewidth]{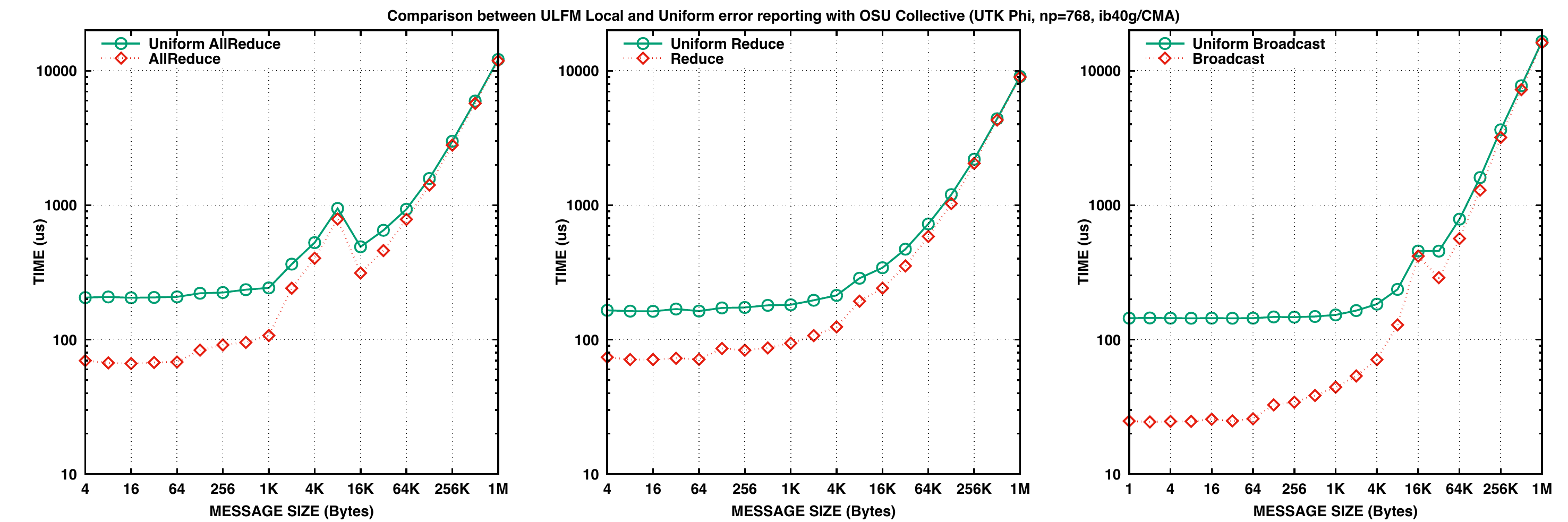}
    \caption{Performance comparison between Local and Uniform collective communication (768 MPI ranks)}
    \label{fig:osuU}
\end{figure*}

\section{Performance Evaluation}\label{sec:perf}

\subsection{Experimental Setup}\label{sec::xpsetup}

We evaluate the performance impact (or lack thereof) of the proposed new APIs and communication modes on the Phi cluster, at the University of Tennessee.
Each node consists of a Xeon X5660 12 core CPU, and is connected through a QDR (40Gb/s) MT26428
Infiniband adapter.
In total there are 66 nodes available.
Failures are injected by having selected application processes raise the SIGKILL signal.

The implementation of the new capabilities forks from the \ulfm integrated in the \ompi main branch (at commit hash 437d70d4).
The assets used to produce the results are available from the public \ulfm testings repository from the tag \emph{ftxs22}~\footnote{https://github.com/ICLDisco/ulfm-testing/tree/ftxs22}.

We use the \ompi components OB1/UCT combination to drive the Infiniband network over the UCX 1.11.2, rdma-core 33.1 drivers.

\subsection{Benchmarks}

We have produced a number of specific benchmarks to help investigate the performance impact of the proposed new operations.
First we have a modified version of the OSU MPI benchmarks (forked from OSU 5.7.1) in which the user can select on the command line if error reporting must be local, group-based, or global, on one hand, and if error uniformity must be local or group based.
Second, we produced a new micro-benchmark that can investigate the use of multiple ISHRINK operations running concurrently.
Third, we produced a new micro-benchmark that mimics a \emph{buddy-checkpointing} application, and employs ISHRINK to overlap the cost of restoring the checkpoint dataset with the cost of reconstructing communicators.
A buddy checkpointing application stores a copy of its dataset on another process (its buddy), so that the lost data can be recovered in the event that this process fails.
In this test, the repaired (shrunk) communicator is not yet available to exchange the dataset during recovery, thus the benchmark employs point-to-point communication between the buddy and the process assigned to take on the work of the failed process.
When the recovery procedure is complete, the shrunk communicator can be used by the recovered processes to resume using collective communication.

\subsection{Effect of in-band versus out-of-band Error Reporting on Micro-Benchmarks}

Fig.~\ref{fig:osuE} presents the results comparing the cost of using the in-band versus out-of-band error detection capabilities, as required to support the local, versus group error reporting modes respectively.
When fault tolerance is ``off'', the out-of-band error detection subsystem in Open MPI is disabled and only in-band errors are produced (from the Infiniband transport).
In contrast, when we enable fault tolerance, and set the communicator on which the test operates in the ``group'' error reporting range, Open MPI is forced to enable the out-of-band error detection subsystem.
We compare the performance in both bandwidth and latency for a number of OSU micro benchmarks, without faults.
The benchmarks are run with a large number of iterations to increase runtime, and thus increase the number of heartbeat events produced during each run (heartbeat is set to every 5s).
One can observe that the performance in both collectives is completely unchanged, and point-to-point microbenchmarks show only a very slight latency degradation (from 2.22$\mu s$ to 2.45$\mu s$) with no bandwidth impact, illustrating that the addition and use of the new reporting modes does not impact performance significantly.

\subsection{Effect of Uniform Mode on Collective Communication Performance}

Fig.~\ref{fig:osuU} presents the results comparing the cost of using the uniform error reporting in a set of collective communication patterns in OSU micro benchmarks.
Time reported in the figure are the ``maximum'' report from OSU, that is, the time to complete the operation at all ranks (the default ``average'' report computes the average of the completion time at each rank, per operation, a measure of how synchronizing the operation is).
The benchmarks are selected to cover one-to-many, many-to-one, and many-to-many communication patterns.
One can observe that in all cases, the addition of the verification step to enforce uniform error reporting has a significant impact on small message latency.
In addition, in one-to-many and many-to-one operations, there is a significant departure in the inter-rank distribution of completion times: when the error reporting is local, some processes (presumably close to the root of the broadcast topology) complete the broadcast operation much more quickly than other processes (presumably leaves); in the uniform mode, the operation becomes synchronizing in a strong sense and all ranks experience a similar wait time for the broadcast operation to complete.
In terms of bandwidth, the difference between the local and uniform modes decreases with message size, because the fixed cost of the verification step can be amortized with the increased cost of the communication itself.

\begin{figure}
    \includegraphics[width=\linewidth]{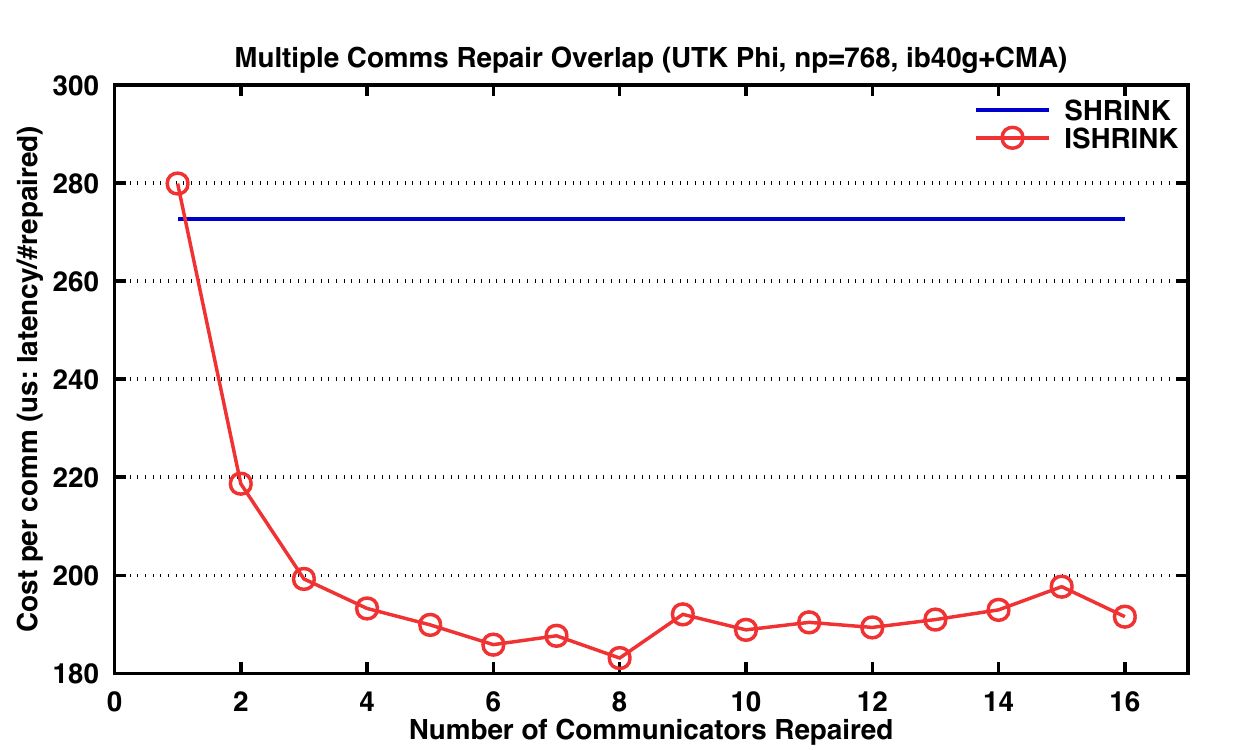}
    \caption{Overlap between multiple concurrent ISHRINK operations (768 MPI ranks)}
    \label{fig:ishrink-multi}
\end{figure}

\begin{figure}
    \includegraphics[width=\linewidth]{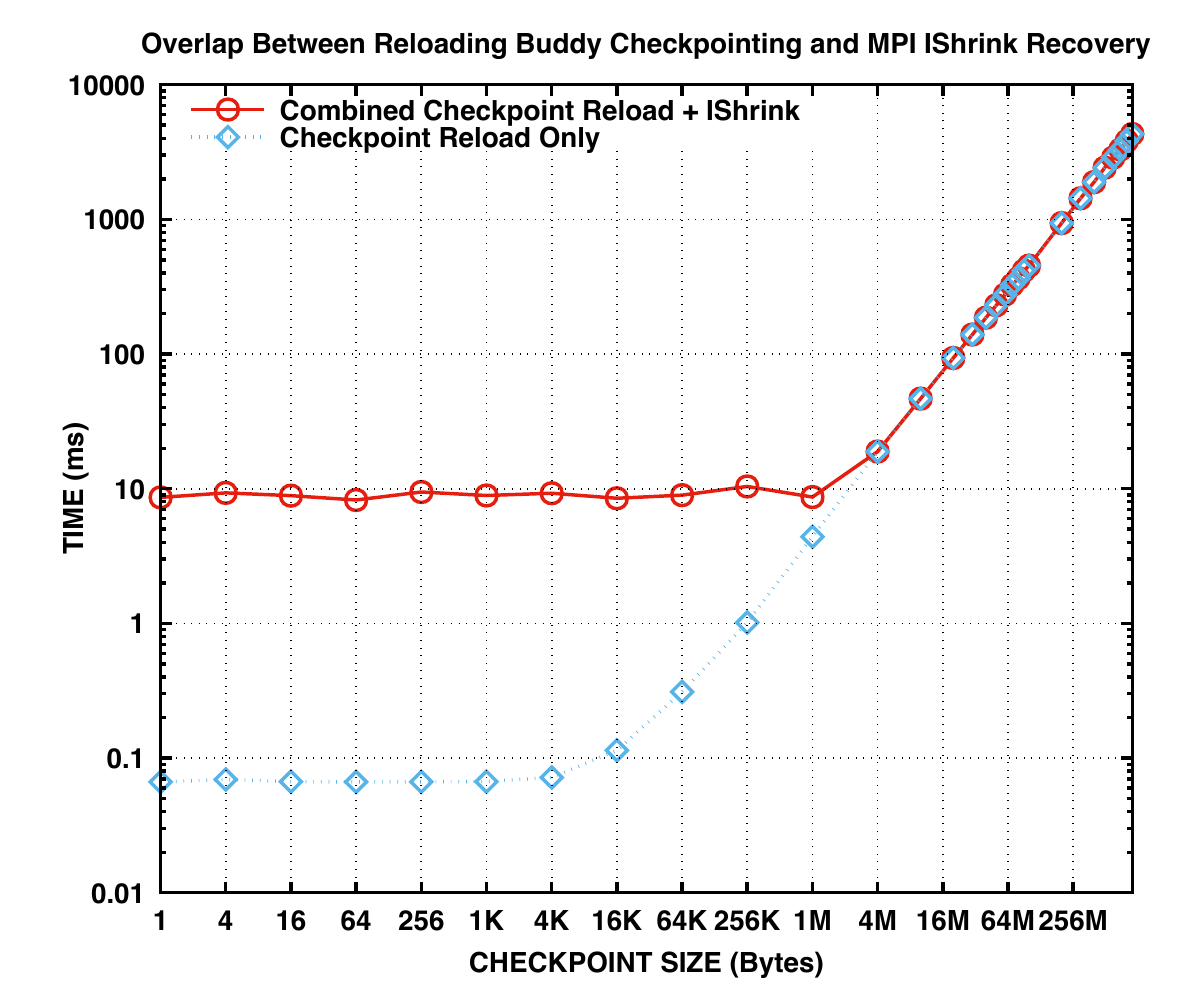}
    \caption{Overlap between ISHRINK and Buddy Checkpointing Reload (768 MPI ranks)}
    \label{fig:ishrink-ckpt}
\end{figure}

\subsection{Concurrent Shrinks Overlap}

Fig.~\ref{fig:ishrink-multi} presents the performance observed when rebuilding multiple communicators after a single rank process fault.
When using the non-blocking shrink operation, multiple communicators can be repaired concurrently.
We present the per communicator cost, that is, the time it takes to repair the set of communicators, divided by the number of communicators.
The first observation is that using the blocking or non-blocking variants of the shrink operation takes approximately the same time to repair a single communicator; we observe only a small overhead from the split between initiation and completion in the non-blocking variant.
When we increase the number of communicators that are repaired simultaneously (a pattern that can only be achieved using the non-blocking shrink variant), we observe that the cost per communicator decreases, up to approximately 8 communicators repaired simultaneously.
This indicates that applications can reduce the time to recovery by overlapping the repair of multiple communicators with concurrent non-blocking shrink calls.

\subsection{Shrink and Checkpoint Reload Overlap}

Fig.~\ref{fig:ishrink-ckpt} presents the time to recovery after a single rank failure when the application overlaps the time to reload a \emph{buddy checkpoint} with the cost of repairing the communicators (as is needed when the application is intent on using collective communication after the recovery completed).
We vary the size of the checkpoint dataset from very small (1 Byte) to large (1GB).
At the scale of our experiments, the cost of performing a single shrink operation is very small (around 800$\mu s$).
The cost of repairing a communicator after a fault (thus including fault detection inside a shrink operation) is 10$ms$.
On this system, the cost of reloading the buddy checkpoint is equal to the cost of performing the shrink communicator repair operation between checkpoint sizes 1MB to 4MB.
Of note, for the 4MB checkpoint, the time to reload a checkpoint is in the same order of magnitude as the cost of a shrink operation, yet the combined time is in-trend hinting that for this checkpoint size, the cost of the shrink is completely overlapped.
The checkpoint sizes for which we achieve overlap without the sheer cost of the checkpoint amortizing the cost of a serial shrink are rather small on this system.
However, larger scale experiments carried in the past with ULFM show that, while the cost of buddy checkpointing is constant with the number of nodes (it varies with the per-node dataset size), the cost of performing a shrink operation increases with scale, which increases the importance of overlapping the shrink cost at scale, even for larger checkpoint loads.


\section{Related Works}\label{sec:related}


Other models of fault tolerance in MPI have also been researched.
A number of projects deploy masking fault tolerance (i.e., failures are automatically handled by the environment, and MPI processes are replaced implicitly from replicates or checkpoints~\cite{2007-1,BHKLC06,cl85,SSBLDHR03}). One detriment to masking fault tolerance is the typically high cost incurred on failure-free operation.
FT-MPI~\cite{fagg04:_ftmpi} presented an earlier attempt at enabling fault management in MPI applications.
FT-MPI automatically repairs the predefined \mpicode{MPI_COMM_WORLD} communicator based on a
replace, shrink, or blank policy. The shrink and replace modes are synchronous, and the approach is global in nature.
MPI Reinit~\cite{doi:10.1002/cpe.4863,doi:10.1177/1094342015623623} proposes a rollback model in which faulty processes are replaced automatically, all processes restart after the MPI initialization, and data restoration remains under the user's control.
By definition this approach is global and is fully synchronous.
FMI~\cite{Sato:2014} introduces a comprehensive model that handles fault tolerance, including
checkpointing the application state, restarting failed processes and automatically reallocating additional nodes.
Compared to \ulfm---which proposes a flexible low-level API that supports a variety of fault tolerance models---these alternatives propose
embracing a monolithic recovery model that supports a single mode of recovery, one that is always operating at a global scope.

For fault tolerance frameworks and languages, multiple frameworks that simplify checkpoint-restart, and user-controlled, in-place, global restart have been implemented on top of \ulfm.
For example,  Fenix~\cite{Gamell:2014,fenix_spec1.0.1} captures errors reported from \ulfm, and rollback the application to the exit of the initialization
function with repaired communicators (in shrink or replace mode).
An explicit data management interface permits saving and restoring application data thereafter.
In NR-MPI~\cite{nrmpi}, failures are detected from the resource manager rather than from MPI operations, and NR-MPI takes care of replacing failed processes (using \ulfm) and triggers data recovery procedures at the application level.
In~\cite{Hassani:2014:CCG:2642769.2642776}, the authors propose an alternative error reporting model based on operational timeouts and employ the \ulfm implementation to prototype their effort.
In the X10 language, an exception-based management of process failures is implemented on top of \ulfm~\cite{x10ft}.
In the Co-Array Fortran language, the concept of ``failed images'' abstracts the detection and elimination of failed processes and is implemented using \ulfm
shrink operations~\cite{cafft}.
In the Local Failure Local Recovery (LFLR)~\cite{lflr} framework, failure management is inserted in the
application through C++ inheritance of recoverable data structures that require protection.
The \ulfm-based implementation captures MPI errors and substitutes spare processes on which
recoverable data is re-instantiated.
In~\cite{DBLP:journals/tjs/LosadaCMG17} the author provided a framework for placing checkpoints at semantic synchronization points.
In~\cite{8514912}, the authors survey different fault tolerance techniques in MPI and their applicability to the Fortran language. In~\cite{Szpindler:2015:EAF:2820083.2820094}, the authors study the cost of re-spawning processes and identify it as one of the largest overheads left standing on the recovery path.
Some of these framework developers have expressed interest in the addition of asynchronous recovery of communicators as an important use-case for their
workflow (e.g., Fenix, LFLR), and this emerging need has motivated the work presented in this paper.

On the application side, in~\cite{Pauli201524}, the authors study a fault tolerant algorithm for the resolution of 2-D stochastic Euler equations of gas dynamics with a monte-carlo method;
in~\cite{NLA:NLA2059}, the authors study numerical recovery strategies for Krylov solvers;
in~\cite{Ahn20151296}, the authors study sender-based message logging deployed at the application level;
in~\cite{Amatya:2017:FTD:3127024.3127037}, the authors study the requirements for deploying deep learning fault tolerance;
and in~\cite{Goddeke:2015:FFM:2843079.2843270}, the authors present an application-based checkpoint compression strategy for finite-element multi-grid algorithms.
Many of these Applications demonstrate usage patterns that recover multiple communicators, and recover small group communicators in isolation from one-another, a good fit for the group and local scoping, as well as non-blocking shrink overlap between communicator repairs.

Nontraditional HPC workloads (e.g., distributed database management systems [DDBMS]~\cite{dbiboip}) and big data analytics on the cloud~\cite{bigdataanalytics,Hoefler:2009:TEM:1612208.1612248,mpihadoop} have also expressed interest in using MPI but had been deterred by its lack of resilience support.
Using \ulfm, they have obtained very impressive performance speedup and demonstrated that a fault-tolerant MPI can clearly serve
some of their needs.
In~\cite{Guo:2015:FTM:2807591.2807617}, the authors evaluated an MPI-based implementation of Hadoop and Map-Reduce with fault tolerance;
in~\cite{sap}, the authors port the SAP database system over \ulfm and demonstrate an impressive speedup when compared to system-level checkpoint and restart; the usage pattern would particularly benefit from the uniform collective mode of operation proposed in this paper.
In~\cite{Zounmevo201343}, the authors consider a wish list of features required to execute cloud compute services over MPI and consider how to match their needs with \ulfm features.

\section{Concluding Remarks}\label{sec:conclusion}

In the past, MPI fault recovery has considered only monolithic
solutions, where either the entire application restarts, or the
management of failures is totally transactional and contingent on the
messaging pattern. In contrast, this work enable new patterns
for fault tolerance---patterns that do not rely on a blocking, serialized,
chain of parallel steps but instead enable merging and overlapping
multiple steps of the recovery.

We have performed an evaluation of the proposed ideas in a practical setting using a mature software environment.
The performance demonstrate that enabling a flexible scoping in error reporting and asynchronous, implicit recovery actions does have the potential to reduce the cost, while at the same time maintaining a level of performance indistinguishable from the non fault-tolerant implementation.

Although our proposal and evaluation is in the
context of MPI, the concepts and designs are expected to carry over to
other programming models (e.g., partitioned global address space
models~\cite{DBLP:conf/openshmem/BouteillerBV16}) beyond a narrow
message-passing classification.
These new capabilities form a new basis upon which
application researchers and system designers can investigate how
asynchronous recovery can be deployed in their own application domain.

\section*{Acknowledgments}
This material is based upon work supported by the National Science Foundation under Grant No. 1664142 SI2-SSI: EVOLVE: Enhancing the Open MPI Software for Next Generation Architectures and Applications.

This research was supported by the Exascale Computing Project (17-SC-20-SC), a collaborative effort of the U.S. Department of Energy Office of Science and the National Nuclear Security Administration.


\bibliographystyle{IEEEtran}
\bibliography{references}


\end{document}